\documentclass[a4paper,12pt]{article}
\usepackage{amsmath}
\usepackage{amsfonts}
\usepackage{amssymb}

\newtheorem{theorem}{Theorem}[section]
\newtheorem{corollary}[theorem]{Corollary}
\newtheorem{lemma}[theorem]{Lemma}

\newtheorem{definition}[theorem]{Definition}

\def\RR{\mathbb{R}}

\def\NN{\mathbb{N}}
\def\CC{\mathbb{C}}
\def\eps{\epsilon}
\newcommand{\Tr}{\mathop{\rm Tr}\,}

\title{Discrete-Time Path Distributions on Hilbert Space\footnote{This article is dedicated to the memory of Erik Thomas}}
\author{M. Beau \& T. C. Dorlas \\Dublin Institute for Advanced Studies \\
School of Theoretical Physics \\ 10 Burlington Road, Dublin 4,
Ireland. }

\begin{document}

\maketitle

\begin{abstract}
We construct a path distribution representing the kinetic part of
the Feynman path integral at discrete times similar to that
defined by Thomas \cite{Thomas}, but on a Hilbert space of paths
rather than a nuclear sequence space. We also consider different
boundary conditions and show that the discrete-time Feynman path
integral is well-defined for suitably smooth potentials.
\end{abstract}

\tableofcontents

\section{Motivation and basic set-up}

\setcounter{equation}{0}
\renewcommand{\theequation}{\arabic{section}.\arabic{equation}} 
\setcounter{section}{1}
\renewcommand{\thesection}{\arabic{section}} 

\subsection{Feynman path integral as a path distribution}

In the Lagrangian formulation of quantum mechanics one defines the
\textit{action} of a particle as an integral of the Lagrangian
over the time duration of the motion:
\begin{eqnarray*}
S(x_f,t_f;x_i,t_i) = \int_{t_i}^{t_f}dt\, L(x(t),\dot{x}(t),t).
\end{eqnarray*}
In general, the Lagrangian $L(x(t),\dot{x}(t),t)$ depends
explicitly on the time, as well as on the position $x(t)$ and the
velocity $\dot{x}(t)$ of the particle. For one-dimensional motion,
the Lagrangian has the form
\begin{eqnarray*}
L(x(t),\dot{x}(t),t)=\frac{m}{2}\dot{x}(t)^2-V(x(t),t)\ ,
\end{eqnarray*}
where the first term is the kinetic energy term and $V(x(t),t)$ is
the external potential. The time-evolution of a wave function
$\Psi(x,t)$ is then given by
\begin{equation} \Psi(x_f,t_f) = \int K(x_f,t_f; x_i,t_i)
\, \Psi(x_i,t_i) dx_i, \end{equation}
where the \textit{propagator}
$K(x_f,t_f; x_i,t_i)$ is given by a \textit{path integral} of the
form
\begin{equation} K(x_f,t_f; x_i,t_i) = \int e^{i S(x_f,t_f;
x_i,t_i)/\hbar}\, {\cal D}[x(t)]. \end{equation} Here ${\cal
D}[x(t)]$ indicates a putative \lq\lq continuous product'' of
Lebesgue measures ${\cal D}[x(t)] = \prod_{t\in (t_i,t_f)} dx(t)$.
(Note that the action $S$ above is a functional of the path
$x(t)$.) It is a formidable mathematical challenge to make sense
of this path-integral concept. Feynman himself interpreted it
loosely as a limit of multidimensional integrals. However, as
Thomas\cite{Thomas} remarks, even the finite-dimensional integrals
are not proper integrals, though they can be defined as improper
integrals. It was already noted by Cameron\cite{Cameron} that the
path integral cannot be interpreted as a complex-valued measure.
In fact, as Thomas \cite{Thomas} and Bijma \cite{Bijma} show, it cannot even be
interpreted as a summable distribution because the summability
order diverges as the number of integrals tends to infinity.

Various alternative approaches have been proposed to interpret the
Feynman path integral as a limit of regularised integrals, e.g.
\cite{CS, Truman, Nelson}. The \lq Euclidean approach' of \lq Wick
rotating' the time in the complex plane has led to the development
of Euclidean quantum field theory, which has been the most
successful way of constructing examples of quantum field theories.
However, this still leaves open the question as to how the path
integral object should be interpreted mathematically. De
Witt-Morette \cite{DWM} has argued that it should be a kind of
distribution, but her approach was formal rather than
constructive. The It\^o-Albeverio-H\o egh Krohn \cite{AHK}
approach was more constructive. They gave a definition of the path
integral as a map from the space of Fourier transforms of bounded
measures to itself and were able to show, using a perturbation
expansion, that this is well-defined for potentials which are also
Fourier transforms of bounded measures. Albeverio and Mazzucchi
\cite{AM} later extended this approach to encompass polynomially
growing potentials. This approach gives a mathematical meaning to
the path-integral expression for the solution of the Schr\"odinger
equation with initial wave function, rather than the propagator. A
different approach in terms of functionals of white noise was
proposed by Hida, Streit et al. \cite{HKPS}. In both approaches,
the space of \lq paths' is rather abstract.

In \cite{Thomas}, Thomas initiated a different approach, with the
aim of defining the path integral as a generalised type of
distribution, in the spirit of De Witt-Morette, 
which he called a \textit{path distribution}. 
In fact, this project is only at the beginning stages. In
\cite{Thomas}, he constructed an analogue of the path integral in
discrete time, where the paths are sequences in a certain nuclear
sequence space. His main idea is to define the path integral as a
derivative of a measure, which we call the Feynman-Thomas measure.
In this paper, we simplify his approach by defining the path
distribution on a space of paths in a Hilbert space instead. This
makes the construction more explicit and the technical details
less demanding.

In the following, we set $m=1$ and $\hbar=1$ for simplicity.
Discretising the action to a finite subdivision
$\sigma=\{t_1,...,t_n \}$ with $0 = t_0 < t_1 < \dots < t_n < T$
and $x=(x_1,..,x_n)\in \RR^n$ we can consider different boundary
conditions. For Dirichlet boundary conditions (DBC) we have
$$x(t=0)=0;\ \ x(t=T)=X_T\ ,$$
and
$$ S(X_T,T;0,0)=\lim_{n\rightarrow\infty} S_n^{(DBC)}(X_T,T;0,0)\ ,$$
where : \begin{eqnarray} && S_n^{(DBC)}\left( x_{n+1}=X_T,
t_{n+1}=T\, ;
\, x_0=0,t_0=0 \right) = \nonumber \\
&& \qquad = \frac{1}{2} \left(\frac{(X_T-x_n)^2}{T-t_n} +
\frac{(x_n-x_{n-1})^2}{t_{n}-t_{n-1}} + \ldots +
\frac{(x_2-x_{1})^2}{t_2-t_{1}} + \frac{x_{1}^2}{t_{1}} \right).
\end{eqnarray} Alternatively, we can impose mixed boundary conditions (MBC):
$$x(t=0)=0;\ \ \dot{x}(t=T)=v_T $$
in which case the action depends on the initial position and the
final velocity $S=S(v_T,T;x_i,t_i)$, so that
$$ S(v_T,T;0,0) = \lim_{n\rightarrow\infty}
S_n^{(MBC)} \left(\frac{x_{n+1}-x_{n}}{t_{n+1}-t_{n}} =
v_T,T;x_0=0,t_0=0 \right),$$ where
\begin{eqnarray} \label{MBCaction}
&& S_n^{(MBC)} \left(
\frac{x_{n+1}-x_{n}}{t_{n+1}-t_{n}}=v_T,t_{n+1}=T\,;\,x_0=0,t_0=0
\right) = \\ &&\quad = \frac{1}{2} \left( v_T^2 (T-t_n) +
\frac{(x_{n}-x_{n-1})^2}{t_{n}-t_{n-1}} + \ldots +
\frac{(x_2-x_{1})^2}{t_2-t_{1}} + \frac{x_{1}^2}{t_{1}} \right)\!.
\end{eqnarray}

The corresponding Feynman distributions are as follows
\begin{eqnarray}
F_{\sigma}^{(DBC)} &=& \frac{1}{\sqrt{2 i \pi (T-t_n)}} \exp
\left[ \frac{i}{2} \left( \frac{(X_T-x_n)^2}{T-t_n} +
\frac{(x_n-x_{n-1})^2}{t_n-t_{n-1}} \right. \right.  \nonumber \\
&& \left. \left. + \ldots + \frac{(x_2-x_1)^2}{t_2-t_1} +
\frac{x_1^2}{t_1} \right) \right] \prod_{j=1}^{n}
\left(\frac{dx_{j}}{\sqrt{2 i \pi (t_{j}-t_{j-1})}} \right)
\end{eqnarray}
and
\begin{eqnarray} \label{FMBC}
F_{\sigma}^{(MBC)} &=& \exp \left[\frac{i}{2} \left(
\frac{(x_n-x_{n-1})^2}{t_n-t_{n-1}} + \ldots +
\frac{(x_2-x_1)^2}{t_2-t_1} + \frac{x_1^2}{t_1}  \right) \right]
\nonumber \\ && \qquad \times  \prod_{j=1}^{n}
\frac{dx_{j}}{\sqrt{2 i \pi (t_j-t_{j-1})}} ,
\end{eqnarray}
for the (MBC) with $v_T=0$.

The Fourier transform of $F_\sigma$ is obtained by computing the
covariance matrix as the inverse of the coefficient matrix. In the
case of mixed boundary conditions this yields
\begin{equation}\label{FTFMBC} \widehat{F}_\sigma^{(MBC)} =
\bigg\langle \exp \left[i \sum_{k=1}^n \xi_k x_k \right], \,
F_\sigma^{(MBC)} \bigg\rangle = e^{-\frac{i}{2} \langle \xi,
K_\sigma \xi\rangle},
\end{equation}
\begin{equation}\label{KijMBC}
\langle \xi, K_\sigma \xi\rangle =
\sum K_{i,j} \xi_i \xi_j \mbox{ with}\ K_{i,j} = \min{(t_i,t_j)}.
\end{equation}
The case of Dirichlet boundary conditions is slightly more
complicated:
\begin{equation} \widehat{F}_\sigma^{(DBC)} =
\bigg\langle \exp \left[i \sum_{k=1}^n \xi_k x_k \right], \,
F_\sigma^{(DBC)} \bigg\rangle = \frac{1}{\sqrt{2 i \pi T}}
e^{-\frac{i}{2} \left( \langle \xi, K^{(T)}_\sigma \xi\rangle +
X_T^2/T \right)},
\end{equation}
\begin{equation} \langle \xi, K^{(T)}_\sigma \xi\rangle =
\sum K^{(T)}_{i,j} \xi_i \xi_j \mbox{ with}\ K^{(T)}_{i,j} =
\min{(t_i,t_j)} \left( 1- \frac{1}{T} \max (t_i,t_j) \right).
\end{equation}

\textbf{Remark.} Notice in particular that $\langle 1\, , \,
F_\sigma^{(MBC)} \rangle = 1$ whereas \\ $\langle 1\, , \,
F_\sigma^{(DBC)} \rangle = \frac{1}{\sqrt{2 i \pi T}}
e^{-\frac{i}{2T} X_T^2}$ in accordance with \cite{FH}.

Both $F_\sigma^{(DBC)}$ and $F_\sigma^{(MBC)}$ are summable
distributions of sum-order $n+1$ by Theorem 3.1 of \cite{Thomas}.

As in \cite{Thomas}, we now change our point of view and fix $t_n
- t_{n-1} = 1$ and seek to define a limiting distribution on a
space of sequences $(x_i)_{i=1}^\infty$ as $n \to \infty$.
Note that in this case $K_{ij}=\min{(i,j)}$ by (\ref{KijMBC}).
Rather than on a nuclear sequence space, however, we will construct a
path distribution on a Hilbert space of sequences.

\subsection{Feynman-Thomas Measure on $\RR^n$}

The main idea of \cite{Thomas} is to define the path \lq integral'
as a path distribution obtained as the derivative of a measure.
Because the order of the distribution is 2 \textit{in each
variable} we need to take 2 derivatives in each variable. We
therefore define the differential operators
\begin{equation} D^{(n)} = \prod_{i=1}^n \left( 1 - \alpha_i^2
\frac{\partial^2}{\partial x_i^2}\right), \end{equation} where the
positive constants are arbitrary. It has a corresponding Green's function
$M_\alpha$ given by
\begin{equation} M_\alpha^{(n)} (x_1,\dots,x_n) = \prod_{i=1}^n
\frac{1}{2\alpha_i} e^{-|x_i|/\alpha_i}, \end{equation} that is,
\begin{equation} \label{Green} D^{(n)} M_\alpha^{(n)} =
\delta(x_1) \dots \delta(x_n). \end{equation}

We can now define the path distribution $F^{(n)}$ given by
(\ref{FMBC}) as the derivative \begin{equation} F^{(n)} = D^{(n)} \mu^{(n)} \end{equation} of a bounded complex-valued measure $\mu^{(n)}$ which according to (\ref{Green}) is given by the convolution product $M^{(n)}_\alpha * F^{(n)}$.
To determine this convolution product, we use the representation
\begin{equation}\label{RepresM}
 \frac{1}{2\alpha} e^{-|x|/\alpha} =
 \int_0^{+\infty} \frac{ds}{\beta}\, e^{-s/\beta}
 \frac{e^{-x^2/2s}}{\sqrt{2 \pi s}},\ \mathrm{with}\ \beta=2\alpha^2\ .
\end{equation}
This can be obtained by Fourier transformation:
\begin{eqnarray*}
  \int_0^{+\infty} \frac{ds}{\beta}\, e^{-s/\beta}
 \frac{e^{-x^2/2s}}{\sqrt{2 \pi s}}&=&
 \int_0^{+\infty} \frac{ds}{\beta}\, e^{-s/\beta}
 \int_{\RR}\frac{dk}{2\pi} e^{i k x} e^{-s k^2/2}
{}\\{}&=&\int_{\RR}\frac{dk}{2\pi\beta}\frac{e^{i k x}}{\beta^{-1}+k^2/2}\ ,
\end{eqnarray*}
which implies (\ref{RepresM}) using the residue theorem.
The representation (\ref{RepresM}) yields an explicit formula for what we may call the
Feynman-Thomas measure on the finite sequence space $\RR^n$:
\begin{definition}\label{DefMu}
The Feynman-Thomas measure $\mu^{(n)}$ on $\RR^n$ is defined by
\begin{eqnarray}\label{measure}
\mu^{(n)}(dx_1\ldots dx_n) &=&  M^{(n)}*F^{(n)}(dx_1 \ldots dx_n)
\nonumber \\ &=& \left(\int_{[0,+\infty)^n} \nu^{(n)}(dS^{(n)}) \,
G_{A^{(n)}} (x_1,\dots,x_n)\right) dx_1\ldots dx_n \nonumber \\
\end{eqnarray}
where we integrate over the variables $s_1,..,s_n$
\begin{equation} \nu^{(n)}(dS^{(n)}) = \prod_{i=1}^n
\frac{1}{\beta_i} e^{-s_i/\beta_i} ds_i \mbox{ with } \beta_i = 2
\alpha_i^2 \ .
\end{equation}
and where $G_{A^{(n)}}$ is the convolution product of $(x_1,..,x_n)\mapsto\prod_{j=1}^n(2\pi s)^{-1/2}e^{-x_j^2/2s}$
with $F^{(n)}(dx_1,..,dx_n)$.
Then, using (\ref{FTFMBC}) and (\ref{RepresM}), the Fourier transform of the complex
Gaussian  $G_{A^{(n)}}(x_1,..,x_n)$ is given by
\begin{equation}\label{FTGA}
\widehat{G}_{A^{(n)}}(\xi_1,..,\xi_n) = e^{-\langle
A^{(n)}\xi^{(n)},\xi^{(n)}\rangle/2}
\end{equation}
where $\xi^{(n)} = (\xi_1,\dots,\xi_n)\in \RR^n$ and where $A^{(n)} =
S^{(n)} + iK^{(n)}$, $S^{(n)}=diag(s_1,\dots,s_n)$ and $K^{(n)}_{i,j} = \min{(i,j)}$.
Notice that $A^{(n)}$ is a complex symmetric matrix with positive real part
which implies that it is invertible \cite{Thomas}.
Hence, by computing the inverse Fourier transform, we get
\begin{equation}\label{GA}
G_{A^{(n)}}(\xi_1,..,\xi_n) = \frac{1}{\sqrt{(2\pi)^n\det{(A^{(n)})}}} e^{-\langle
(A^{(n)})^{-1} x^{(n)},x^{(n)}\rangle/2} \ ,
\end{equation}
\end{definition}

It was shown in \cite{Thomas} that $\mu^{(n)}$ is a bounded
complex-valued measure on $\RR^n$. The aim of this work is to
prove that there exists a measure $\mu$ on an infinite dimensional
Hilbert space of paths, given by the projective limit of the
finite-dimensional measures $\mu^{(n)}$, i.e.
$\mu=\varprojlim\mu^{(n)}$.

\section{Hilbert spaces of paths}

\subsection{Regularized-$l^2$ spaces}

We introduce a family of Hilbert spaces of sequences labelled by a
real parameter $\gamma$:
\begin{equation}
l^2_\gamma = \{(\xi_i)_{i=1}^{\infty} \in \RR^\infty |\,
\sum_{i=1}^{\infty} i^{\gamma}\,\xi_i^2 <+\infty\}.
\end{equation}
This is a Hilbert space with inner product given by
\begin{equation}
(\xi,\zeta)_{\gamma} = \sum_{i=1}^\infty \xi_i\, \zeta_i\,
i^\gamma \end{equation} (Notice that obviously $l^2_0=l^2$.)

We have the obvious lemmas
\begin{lemma}
The set of vectors $\{e_i^{(\gamma)}\}_{i=1}^{\infty}$, given by
the sequences
$$ \left(e_i^{(\gamma)}\right)_j = \delta_{i,j}
j^{-\gamma/2}, $$ is an orthonormal basis of the Hilbert space
$l^2_\gamma$.
\end{lemma}

and

\begin{lemma}
The Hilbert spaces $ l^2_\gamma $ and  $l^2_{-\gamma}$ are dual
w.r.t. the duality bracket
$$ \langle \xi\, , \zeta' \rangle = \sum_{i=1}^\infty
\xi_i\,\zeta'_i, $$ where $\xi = (\xi_i)_{i=1}^\infty \in
l^2_\gamma$ and $\zeta' = (\zeta'_i)_{i=1}^\infty \in
l^2_{-\gamma}$.
\end{lemma}

We shall construct the Feynman-Thomas measure $\mu$ on a space
$l^2_{-\gamma}$ for a $\gamma > 0$ large enough. The advantage of the Hilbert space approach is that we can use the following
theorem due to V. Sazonov for the existence of the projective limit, the proof of which is quite simple: see the Appendix and \cite{Sazonov}.

\begin{theorem}[V. Sazonov] \label{SazonovThm} Let $(\mu^{(N)})_{N \in \NN}$ be a
projective system of bounded measures on the dual ${\cal H}'$ of a
separable Hilbert space $\cal H$, i.e. there is an orthonormal
basis $\{e_i\}_{i=1}^\infty$ of ${\cal H}$ with dual basis
$\{e'_i\}_{i=1}^\infty$ such that $\mu^{(N)}$ is a bounded (in
general complex-valued) measure on the span of
$\{e'_1,\dots,e'_N\}$, such that for $M > N$, $\pi'_N(\mu^{(M)}) =
\mu^{(N)}$, where $\pi'_N$ is the projection onto the span of
$\{e'_1,\dots,e'_N\}$. Assume that there exist positive measures
$\nu_N$ such that $|\mu^{(N)}| \leq \nu_N$ and which are uniformly
bounded:
$$ \sup_{N \in \NN} ||\nu_N|| < +\infty, $$ and such that the
Fourier transforms $\Phi_N: {\cal H} \to \CC$ given by
$$ \Phi_N(\xi) = \int e^{i \langle \pi_N(\xi),\, x \rangle}
\nu_N(dx), $$ (where $\pi_N$ is the projection on the span of
$\{e_1,\dots,e_N\}$) are equicontinuous at $\xi = 0$ in the
Sazonov topology, i.e. for all $\eps > 0$ there exists a
Hilbert-Schmidt map $u \in {\cal B}({\cal H})$ such that
$$ ||u\,\xi|| \leq 1 \implies |\Phi_N(\xi) - \Phi_N(0)| \leq \eps
\quad \forall N \in \NN. $$ Then there exists a unique bounded
Radon measure $\mu$ on ${\cal H}'_\sigma$, where the subscript
$\sigma$ denotes the weak topology, such that $\pi'_N(\mu) =
\mu^{(N)}$ for all $N \in \NN$.
\end{theorem}

To determine the projective limit of the complex-valued measures
$\mu^{(n)}$ above, we apply this theorem to auxiliary positive
measures which dominate $|\mu^{(n)}|$.

\subsection{Construction of auxiliary measures on $\RR^n$}

We want to construct an auxiliary measure $\mu_{aux}$ to give a majorisation of the modulus of the Feynman-Thomas measure $\mu$, see Definition \ref{DefMu}. Indeed, if we can prove that this auxiliary measure is strongly concentrated on an Hilbert space
$l^2_{-\gamma}$ for some $\gamma > 0$ (i.e. defines a Radon measure $\mu_{aux}$ on this space; this is the case if its total mass is concentrated on a compact set up to arbitrary $\eps > 0$: see e.g. \cite{Thomas2}), then it follows that $\mu$ is also strongly concentrated on $l^2_{-\gamma}$ since $|\mu|\leq \mu_{aux}$. (We remark that the covariance $K = \lim_{n \to \infty} K^{(n)}$ must then be considered as a map $ K:\, l^2_\gamma \to l^2_{-\gamma}$ with kernel $K_{i,j} = \min (i,j)$, so that  $\langle \xi\, ,\, K\,\xi \rangle = \sum_{i,j} K_{i,j} \xi_i\,\xi_j$.) The auxiliary
measure $\mu_{aux}$ will be the projective limit of the measures
$\mu^{(n)}_{aux}$ given by
\begin{equation}\label{auxmeas}
\mu_{aux}^{(n)}(dx_1\ldots dx_n) = \int_{\RR^n_+}
|G_{A^{(n)}}(x_1,..,x_n)| \,\nu^{(n)}(dS^{(n)})\, dx_1\ldots dx_n
\end{equation}
where $G_{A^{(n}}(x_1,..,x_n)$ and $\nu(dS^{(n)})$ are defined in Definition \ref{DefMu}.\\
The Fourier transform with respect to $x$ of the auxiliary measure
is defined by:
\begin{equation}\label{DefFTauxmeas}
\Phi_n(\xi) = \int \mu_{aux}^{(n)}(dx_1\ldots dx_n)\, e^{i \langle
x^{(n)}, \xi^{(n)}\rangle}
\end{equation}

The aim is now to show that the measures $\mu^{(n)}_{aux}$ on
$l^2_{-\gamma}$ satisfy the conditions of Sazonov's theorem. Then
it follows that the projective limit
$$ \mu = \varprojlim \mu^{(n)} $$ exists on $l^2_{-\gamma}$
w.r.t. the weak topology.

Evaluating (\ref{auxmeas}) using (\ref{GA}) we have
\begin{eqnarray}\label{Computeauxmeas}
\mu_{aux}^{(n)}(dx_1\ldots dx_n) &=& \int_{\RR^n_+}
\nu^{(n)}(dS^{(n)}) \exp \left[-\frac{1}{2} \langle x^{(n)},\Re e
\left(({A}^{(n)})^{-1}\right) x^{(n)}\rangle \right] \nonumber
\\ && \qquad \times \frac{dx_1 \dots dx_n}{(2\pi)^{n/2} \left|
\sqrt{\det(S^{(n)}+iK^{(n)})} \right|}
\end{eqnarray}
where $x^{(n)}=(x_1,\dots,x_n)$. To compute the Fourier transform,
we need to determine $B^{(n)} = \left( \Re e(A^{(n)})^{-1}
\right)^{-1}$. Omitting the superscripts for simplicity, we have
$$ A^{-1} = S^{-1/2} \left( I + i S^{-1/2} K S^{-1/2} \right)^{-1}
S^{-1/2}. $$ With $C = S^{-1/2} K S^{-1/2}$,
$$ (I + i C)^{-1} = (I + C^2)^{-1} (I - iC) $$ and since $C$
is real,
\begin{eqnarray*} (\Re e(A)^{-1})^{-1} &=& S^{1/2} (\Re e(I+i C)^{-1})^{-1}
S^{1/2} \\ &=&  S^{1/2} (I+C^2) S^{1/2} = S + K S^{-1} K. \end{eqnarray*}
Thus
\begin{equation} B^{(n)} = \left( \Re e(A^{(n)})^{-1}
\right)^{-1} = S^{(n)} + K^{(n)} (S^{(n)})^{-1} K^{(n)}\ ,
\end{equation}
which is a positive definite symmetric matrix.
We also compute $$ \int |G_{A^{(n)}}(x)|\,  d^nx =
\frac{\sqrt{\det B^{(n)}}}{\sqrt{|\det A^{(n)}|}}. $$ Using $$
|\det A| = (\det S) |\det (I + iC)| = (\det S) \det (I + C^2)
|\det(I-iC)|^{-1} $$ and $$ \det B = (\det S) \det (I+C^2) $$ we
have \begin{eqnarray} \label{normbnd} \int |G_{A^{(n)}}(x)|\, d^nx
= \sqrt{|\det (I - i (S^{(n)})^{-1/2} K^{(n)} (S^{(n)})^{-1/2})|}. \end{eqnarray}
The Fourier transform with respect to the sequence $x^{(n)}$ is then 
(see (\ref{auxmeas}), (\ref{DefFTauxmeas}), (\ref{Computeauxmeas}), (\ref{normbnd})) :
\begin{equation} \label{FTauxmeas}
\Phi_n(\xi) =  \int_{\RR^n_{+}}\nu(d{S^{(n)}}) e^{-\langle \xi^{(n)},{B}^{(n)}
\xi^{(n)}\rangle/2} \sqrt{|\det(1-i(S^{(n)})^{-1} K^{(n)})|} ,
\end{equation}
where we integrate over the variables $s_1,..,s_n$
and where $\xi^{(n)}=(\xi_1,..\xi_n)$ and where $B^{(n)} = S^{(n)} + \Gamma^{(n)}$,
with $\Gamma^{(n)} = K^{(n)} (S^{(n)})^{-1} K^{(n)}$. In the following two subsections we verify that the conditions for Theorem~\ref{SazonovThm} are satisfied on the Hilbert space ${\cal H}' = l^2_{-\gamma}$ for $\gamma > 0$ large enough, i.e. that $||\mu_{aux}^{(n)}||$ is uniformly bounded, and that $\Phi_n(\xi)$ is equicontinuous w.r.t. the Sazonov topology on ${\cal H} = l^2_\gamma$. For the latter it suffices if the quadratic form $\xi \mapsto \langle \xi^{(n)},\, B^{(n)} \xi^{(n)} \rangle$ is equicontinuous in the Sazonov topology on $l^2_\gamma$ for a.e. $S$.

Defining the map $B = S + \Gamma$, with $\Gamma = K S^{-1} K$, which is the inverse of the real part of the inverse of $A$, i.e. $B = \left(\Re e(A^{-1})\right)^{-1}$ as a map: $l^2_\gamma \rightarrow l^2_{-\gamma}$, the Fourier transform of the limiting auxiliary measure ${\mu_{aux}}$ on $l^2_{-\gamma}$ will be given by
\begin{equation}
\Phi(\xi) = \int \nu(d{S}) e^{-\langle \xi, B \xi\rangle/2}
\sqrt{|\det(1-i{S}^{-1}{K})|},\quad \xi \in l^2_\gamma.
\end{equation}

\subsection{Uniform boundedness}

It is clear from (\ref{auxmeas}) and (\ref{normbnd}) that the norm of the measure
$\mu^{(n)}_{aux}$ is given by
\begin{equation} \label{auxnorm} || \mu^{(n)}_{aux} || = \int \nu(dS^{(n)})
\sqrt{|\det(1-i(S^{(n)})^{-1}{K^{(n)}})|}. \end{equation}

\begin{lemma} \label{Det}
We have $$ \sup_{n \in \NN} || \mu_{aux}^{(n)} || < +\infty,\ $$ if the following
condition holds
$$ \sup_{n \in \NN} \sum_{i=1}^n \kappa_i^{(n)} < +\infty $$ where
\begin{equation} \kappa_i^{(n)} = \sqrt{ \sum_{j=1}^n
\frac{|K_{i,j}|^2}{\beta_i \beta_j}}. \end{equation}
\end{lemma}

\textit{Proof:} We omit the superscripts $n$ as before. Define
$\tilde{K}_{i,j} = \frac{K_{i,j}}{\sqrt{\beta_i \beta_j}} $ and
$\tilde{s}_i = s_i/\beta_i$. Then  $$ \kappa_i =\sqrt{
\sum_{j=1}^n |\tilde{K}_{i,j}|^2}. $$ Rescaling, we have
\begin{eqnarray*}
\int && \nu(dS^{(n)}) \sqrt{|\det\left(I - i({S}^{-1/2} {K}
S^{-1/2}\right)|} = \\ && = \int_{\RR^n_+} d^n \tilde{s}\,
e^{-(\tilde{s}_1 + \dots + \tilde{s}_n)} \sqrt{|\det\left(I -
i(\tilde{S}^{-1/2} \tilde{K} \tilde{S}^{-1/2} \right)|} \\ && =
\int_{\RR^n_+} d^n \tilde{s}\, e^{-(\tilde{s}_1 + \dots +
\tilde{s}_n)} \sqrt{|\det\left(I - i(\tilde{S}^{-1} \tilde{K}
\right)|}. \end{eqnarray*}

This can be estimated as in \cite{Thomas} by means of the Hadamard inequality (see e.g. \cite{Bell})
\begin{equation} |\det(A^{(n)})| \leq \prod_{i=1}^n
\sqrt{\sum_{j=1}^n |A_{i,j}|^2}. \end{equation} It follows that (omitting the tilde on $s$)
\begin{equation} \int_{\RR^n_+} d^ns\, e^{-(s_1+\dots + s_n)} \sqrt{|\det (I-i
S^{-1} \tilde{K})|} \leq \prod_{i=1}^n \int_0^\infty ds\, e^{-s}
(1+\kappa_i^2 s^{-2})^{1/4}. \end{equation} Using
$$ (1+x)^{1/4} \leq 1 + x^{1/4} $$ for $s \leq \kappa_i$ and
$$ (1+x)^{1/4} \leq 1 + x/4 $$ for $s > \kappa_i$, we obtain
\begin{eqnarray*}
\int_0^\infty ds\,e^{-s} (1+\kappa_i^2 s^{-2})^{1/4}
&\leq& 1+k_i^{1/2}\int_0^{k_i}e^{-s}s^{-1/2}ds+\frac{1}{4}k_i^2\int_{k_i}^{+\infty}e^{-s}s^{-2}ds
{}\\{}&\leq& 1 + \frac{9}{4} \kappa_i\ ,
\end{eqnarray*}
since $e^{-s}\leq 1$.
Therefore
\begin{equation} \int \nu(dS^{(n)}) \sqrt{|\det\left(I - i({S}^{-1/2} {K}
S^{-1/2}\right)|} \leq \prod_{i=1}^n \left( 1 + \frac{9}{4}
\kappa_i \right) \leq \exp \left[ \frac{9}{4} \sum_{i=1}^n
\kappa_i \right].
\end{equation}
\phantom{xx} \hfill $\square$

\begin{corollary}\label{CorollaryDet} Set $K_{i,j} = i \wedge j$ and assume $\beta_i =
c i^\delta$ for come $c > 0$ and $\delta > 0$. Then
$$ \sup_{n \in \NN} || \mu^{(n)}_{aux} || < + \infty
\mbox{ if } \delta > \frac{5}{2}.  $$ \end{corollary}

\textit{Proof.}

\begin{eqnarray} \label{kappaest}
(\kappa_i^{(n)})^2 &=& \sum_{j=1}^n |\widetilde{K}_{ij}|^2 = c^{-2}
\sum_{j=1}^n \frac{(i \wedge j)^2}{i^{\delta} j^{\delta}} \\ &=&
c^{-2} i^{-\delta} \sum_{j=1}^i j^{2-\delta} + c^{-2} i^{2-\delta}
\sum_{j=i+1}^n j^{-\delta}
\end{eqnarray}

We use the following estimates:
\begin{eqnarray*}
&&\sum_{j=1}^i j^{2-\delta} \leqslant 1+\int_{1}^i dz
z^{2-\delta}=\frac{2-\delta}{3-\delta}+\frac{i^{3-\delta}}{3-\delta}
{}\\{}&&\sum_{j=i+1}^\infty j^{-\delta}\leqslant \int_{i}^\infty
dz z^{-\delta}=\frac{i^{1-\delta}}{\delta-1}
\end{eqnarray*}
with the condition $\delta>2$ (and $\delta \neq 3$).

If $\delta <3$ then it follows that the both terms in
(\ref{kappaest}) behave like $i^{3-2 \delta}$ and hence $\sup_{n
\in \NN} \sum_{i=1}^n \kappa_i^{(n)} < +\infty$ if
$\sum_{i=1}^\infty i^{\frac{3}{2} -\delta} <+\infty$, i.e. $\delta
> 5/2$. If $\delta \geq 3$ the first term dominates and behaves
like $i^{-\delta}$ (or $i^{-3} \ln i$) and the sum $\sum_{i=1}^n
\kappa_i^{(n)}$ is also bounded.
$\square$ \\

\subsection{Equicontinuity of the quadratic forms}

It remains to determine when the quadratic form $\langle
B\xi,\xi\rangle$ is continuous in the Sazonov topology. Since
$$|\langle B\xi,\xi\rangle|\leq |\langle \xi, S \xi\rangle|
+|\langle \xi, \Gamma \xi\rangle|,$$ it suffices to find two
Hilbert-Schmidt maps $u_S$ and $u_\Gamma$ such that:
$$|\langle \xi, S\xi\rangle|\leq||u_S\xi||^2,$$
$$|\langle \xi, \Gamma \xi\rangle|\leq ||u_\Gamma\xi||^2. $$ Here we
note that by unitary equivalence, the image Hilbert space is
arbitrary. We construct the maps $u_R, u_\Gamma: l^2_\gamma \to
l^2$.

\begin{lemma} \label{ContS}
The quadratic form $\langle {S} \xi,\xi\rangle$, $\xi\in l^2_\gamma$, for some $\gamma\geqslant 0$,
is continuous in the sense of Sazonov topology for $\nu-$almost every ${S}=({s}_i)_{i\geq 1}$ if
\begin{eqnarray}
\sum_i \frac{{\beta}_i}{i^{\gamma}} < +\infty. \label{Condition1}
\end{eqnarray}
\end{lemma}

Proof:\\

Let ${S}=({s}_i)_{i\geq 1}$. Then, $$|\langle S\xi,\xi
\rangle|=|\langle \sqrt{S}\xi,\sqrt{S}\xi \rangle| =
||\sqrt{S}\xi||_{l^2}^2\ ,$$ since $S$ is diagonal and positive.
We therefore choose $u_S = \sqrt{S}$ and obtain
\begin{eqnarray*}
||u_S||_{HS}^2 &=& ||\sqrt{S}||_{2,\gamma}^2 = \sum_{i=1}^\infty
|| u_S e^{(\gamma)}_i ||^2 \\ &=& \sum_{i=1}^\infty || \sqrt{s_i}
i^{-\gamma/2} e_i ||^2 = \sum_{i=1}^\infty \frac{s_i}{i^\gamma}.
\end{eqnarray*}
This converges for $\nu-$almost every ${S}$ if :
\begin{eqnarray*}
\sum_{i=1}^\infty \int\frac{{s}_i}{i^{\gamma}} \nu(d{S}) =
\sum_{i=1}^\infty \int_0^\infty \frac{d{s}_i}{{\beta}_i}
\frac{{s}_i}{i^{\gamma}} e^{-{s}_i/{\beta}_i} = \sum_{i=1}^\infty
\frac{{\beta}_i}{i^{\gamma}} <+\infty.
\end{eqnarray*}

\phantom{xx} \hfill $\square$ \\

\begin{lemma}\label{ContH}
The quadratic form $\langle \xi, \Gamma \xi\rangle$, $\xi\in
l^2_\gamma$ for some $\gamma\geqslant 0$ is continuous w.r.t. the Sazonov topology for $\nu-$almost every ${S}$ if
\begin{equation}
\sum_{i=1}^\infty \sqrt{ \sum_{j=1}^\infty
\frac{|K_{i,j}|^2}{\beta_i j^\gamma}} < +\infty.
\label{Condition2}
\end{equation}
\end{lemma}

\textit{Proof.}
Since $\Gamma = K S^{-1} K$, we have $$ |\langle \xi, \Gamma \xi
\rangle| = |\langle S^{-1/2} K \xi, S^{-1/2} K \xi \rangle| =
||S^{-1/2}K\xi||^2\ $$ because $S$ is diagonal and positive and
$K$ is symmetric. Therefore we choose $u_\Gamma = S^{-1/2} K$ and
have the following condition:
\begin{equation}
||u_\Gamma||_{HS}^2 = ||S^{-1/2} K||_{2,\gamma}^2 < +\infty\
\mathrm{for\ \nu-a.e. {S}}
\end{equation}
We compute the Hilbert-Schmidt norm:
\begin{eqnarray*} ||u_\Gamma ||^2_{HS} &=& \sum_{i=1}^\infty ||
S^{-1/2} K e^{(\gamma)}_i||^2 \\ &=& \sum_{i=1}^\infty ||
\sum_{j,k=1}^\infty s_j^{-1/2} K_{j,k} k^{-\gamma/2} \delta_{ik}
e_j ||^2 \\ &=& \sum_{i=1}^\infty \sum_{j=1}^\infty \frac{1}{s_j}
|K_{j,i}|^2 i^{-\gamma}. \end{eqnarray*} Since
$$ \sum_{i=1}^\infty a_i < +\infty \implies \sum_{i=1}^\infty
a_i^2 < +\infty $$ it suffices if
$$ \sum_{i=1}^\infty \frac{1}{\sqrt{s_i}} \sqrt{\sum_{j=1}^\infty
\frac{|K_{i,j}|^2}{j^\gamma}} < +\infty. $$ Now,
$$ \int_0^\infty \frac{1}{\sqrt{s}} e^{-s/\beta} \frac{ds}{\beta} =
\sqrt{\frac{\pi}{\beta}} $$ so the condition (\ref{Condition2})
follows.
$\square$ \\

\section{Existence of the Feynman-Thomas measure on $l^2_{-\gamma}$}

\begin{theorem}\label{Thm1}
Consider the map $K:l^{2}_{\gamma}\rightarrow l^2_{-\gamma}$ with
$K_{i,j} = i \wedge j$, and assume $\gamma > \frac{7}{2}$. Then
 there exists a unique path distribution $F_K$ on $l^2_{-\gamma}$ such that
$\widehat{F}_{{K}}(\xi)=e^{-i\langle{K}\xi,\xi\rangle/2}$ given by
${F}_K={D}{\mu}$ where ${D}=\prod_{i=1}^\infty
\left(1-\frac{{\beta}_i}{2} \frac{\partial^2}{\partial x_i^2}
\right)$ and where ${\mu}$ is a bounded Radon measure strongly
concentrated on $l^2_{-\gamma}$ w.r.t. the weak topology.
\end{theorem}

\textit{Proof.} It suffices to prove that the auxiliary measures satisfy the conditions of Sazonov's theorem. By the above lemmas, it suffices if the following conditions hold:
$$ \beta_i = i^\delta \mbox{ with } \delta > \frac{5}{2}; $$
$$ \sum_{i=1}^\infty \frac{\beta_i}{i^\gamma} < +\infty; $$
and $$ \sum_{i=1}^\infty \sqrt{\sum_{j=1}^\infty
\frac{|K_{i,j}|^2}{\beta_i j^\gamma}} < +\infty. $$ The first two conditions hold if $\gamma > \frac{7}{2}$ and the proof of
the the Corollary \ref{CorollaryDet} then shows that the last condition is also fulfilled. $\square$

\begin{corollary} Suppose that the potential $V: \RR \to \RR$ belongs to ${\cal E}^{(2)}(\RR)$,
i.e. it is twice continuously differentiable with bounded first
and second derivatives. Moreover, let $(\lambda_j)_{j=1}^\infty$
be a sequence of positive constants such that $ \sum_{j=1}^\infty
\beta_j \lambda_j < +\infty, $ where the constants $\beta_j$
satisfy the conditions of the above lemmas, in particular if
$\beta_j = c\,i^\delta$ with $\delta > 5/2$. Then the Feynman \lq
path integral\rq
$$ \bigg\langle \exp \left[ -i \sum_{j=1}^\infty \lambda_j
V(x_j) \right], F \bigg\rangle $$ exists.
\end{corollary}

\textit{Proof.} This follows from the theorem since
$$ \bigg\langle \exp \left[ -i \sum_{j=1}^\infty \lambda_j V(x_j) \right], F \bigg\rangle =
\bigg\langle D \exp \left[ -i \sum_{j=1}^\infty \lambda_j V(x_j)
\right], \mu \bigg\rangle $$ where $\mu$ is the Feynman-Thomas
measure. It therefore suffices if \\ $D \exp \left[ -i
\sum_{j=1}^\infty \lambda_j V(x_j) \right] $ is bounded. But
\begin{eqnarray*} && D \exp \left[ -i \sum_{j=1}^\infty \lambda_j V(x_j) \right] = \\ && \quad
= \prod_{j=1}^\infty \left\{ 1 + \frac{1}{2} \beta_j \left( i
\lambda_j V''(x_j) + \lambda_j^2 (V'(x_j))^2 \right) \right\} \exp
\left[ -i \sum_{j=1}^\infty \lambda_j V(x_j) \right].
\end{eqnarray*}
and $\sum_{j=1}^\infty\beta_j\lambda_j^2<\infty$
since $\sum_{j=1}^\infty\beta_j\lambda_j<\infty$ implies that $\lambda_j\rightarrow0$ as $j\rightarrow\infty$.
\phantom{xx} \hfill $\square$

\textbf{Remark.} In particular, one can take $\lambda_j =
e^{-\eps\, j}$ for small $\eps > 0$. This is quite common
procedure in scattering theory, known as \lq adiabatically
switching off' the potential.

\section{Concluding remarks}

We have defined the Feynman \lq path integral' with the initial condition $x_0 = 0$ at $t=0$. It is
straightforward to modify this definition to allow for a general boundary condition
$x(t) = x_k$ at $t=k$ for an arbitrary integer $k$. Formally, one then has
\begin{equation} \label{Feynmanxk}
F_n^{(MBC)} = \exp \left[\frac{i}{2} \sum_{n=k+1}^\infty
(x_n-x_{n-1})^2 \right] \prod_{i=k+1}^{\infty}
\left(\frac{dx_{i}}{\sqrt{2 i \pi}} \right).
\end{equation}
Denoting \begin{equation} \Psi_k(x_k) = \bigg\langle \exp \left[
-i \sum_{j=k}^\infty V(x_j) \lambda_j \right],\, F_n^{(MBC)}
\bigg\rangle\ , \end{equation} $\Psi_k$ plays the role of a wave
function at time $k$. There is then an obvious recursion relation:
\begin{equation} \label{opeqn} \Psi_k(x_k) = \int \exp \left[ \frac{i}{2}
(x_n-x_{n-1})^2 - i V(x_k)\lambda_k \right] \Psi_{k+1}(x_{k+1})
\frac{dx_{k+1}}{\sqrt{2 i \pi}}. \end{equation} This equation is
the analogue of the integrated Schr\"odinger equation in the
negative-time direction, i.e. $\Psi_t = e^{i (t'-t) H} \Psi_{t'}$
($t' > t$). (It might therefore have been better to define the
Feynman path integral from $-\infty$ to $k$ instead. This would
represent an incoming wave from $t=-\infty$ to the present.) Note
that the integral kernel in (\ref{opeqn}) defines an operator on
the space ${\cal E}^{(\infty)}(\RR)$ of infinitely differentiable
functions with bounded derivatives (if $V$ has bounded first and
second derivatives). This follows easily by integration by parts,
which is the essence of the distributional approach.

Note that in the  Albeverio - H\o egh Krohn approach they assume
that $V$ is the Fourier transform of a measure, and expand $e^{-i
\int V(x(t)) dt}$. Assuming that $\Psi_{k+1}$ is also the Fourier
transform of a measure, i.e.
$$ V(x) = \int e^{ixy} \nu(dy) \mbox{ and } \Psi_{k+1}(x) = \int
e^{ixy} \mu_{k+1}(dy), $$ we can do the same here:
\begin{eqnarray*} \Psi_k(x_k) &=& \sum_{n=0}^\infty
\frac{(-i)^n}{n!} \int \nu(dy_1) \dots \int \nu(dy_n)
e^{i(y_1+\dots +y_n)x_k} \\ && \quad \times \int \mu_{k+1}(dy)
\int \frac{dx_{k+1}}{\sqrt{2 i \pi}} e^{iyx_{k+1}}
e^{\frac{i}{2}(x_{k+1}-x_k)^2} \\ &=& \sum_{n=0}^\infty
\frac{(-i)^n}{n!} \int \nu(dy_1) \dots \int \nu(dy_n)
e^{i(y_1+\dots +y_n)x_k} \\ && \qquad \qquad \qquad \qquad \times
\int \mu_{k+1}(dy) e^{-\frac{i}{2} y^2 + iyx_{k}} \\ &=& \int
\mu_k(dy) e^{ix_ky},
\end{eqnarray*} where
\begin{equation*} \langle f,\,\mu_k\rangle = \sum_{n=0}^\infty
\frac{(-i)^n}{n!} \int \nu(dy_1) \dots \int \nu(dy_n) \int
\mu_{k+1}(dy)  e^{-\frac{i}{2} y^2} f(y_1 + \dots + y_n + y)
\end{equation*} defines a bounded measure.

It is also of interest to consider the more general boundary
condition at $T \to +\infty$. Taking $x_0$ arbitrary, we define
the \textit{classical path} $\overline{x_i} = x_0 + v\,i$, where
$v = \lim_{T \to +\infty} v_T$ is the limiting velocity. Replacing
$x_i$ by  $x_i + \overline{x_i}$ in the MBC action
(\ref{MBCaction}) it becomes
\begin{eqnarray*} S_n^{(MBC)} &=& \frac{i}{2} \left( v^2(T-t_n) +
\sum_{i=1}^n \frac{(x_i + \overline{x}_i -
(x_{i-1}-\overline{x}_{i-1}))^2}{t_i-t_{i-1}} \right) \\ &=&
\frac{i}{2} \sum_{i=1}^n \frac{(x_i-x_{i-1})^2}{t_i-t_{i-1}} +
\frac{i}{2} v^2 T + i v(x_n-x_0). \end{eqnarray*} The second term
on the right-hand side corresponds to the kinetic energy of a
particle with velocity $v$. The factor $e^{ivx_n}$ represents an
outgoing wave with this velocity and $e^{\frac{i}{2} v^2 T}
e^{ivx_n} $ its free evolution. One defines the kernel of the
(adjoint) \textit{wave operator} $(\Omega^-)^*$ at momentum
$k_{out} = v$ (remember that $\hbar =1$ and $m=1$ so that $v =
\frac{\hbar k}{m} = k$) by omitting these factors and then taking
$n \to \infty$. In the discrete-time case we obtain
\begin{eqnarray*} (\Omega^-)^*(k_{out},x_0) &=& \int {\cal D}[x(t)]
\exp \left[\frac{i}{2} \sum_{j=1}^\infty (x_j-x_{j-1})^2\right] \\
&& \qquad \times \exp \left[ - i \sum_{j=1}^\infty V(x_j + x_0 +
k_{out}\,j) \lambda_j - i k_{out} x_0 \right]
\\ &=& \bigg\langle \exp \left[- i
\sum_{j=1}^\infty V(x_j + x_0 + k_{out} j) \lambda_j - i k_{out}
x_0 \right],\, F \bigg\rangle. \end{eqnarray*}

In scattering theory, one usually considers a time interval which
is unbounded in both directions. One then needs nontrivial
boundary conditions at both ends. We put
\begin{equation} \label{Feynmanscatter}
F_n^{(sc)} = \exp \left[\frac{i}{2}  \sum_{j=-n+1}^n
(x_j-x_{j-1})^2 \right] \prod_{j=-n+1}^{n} \frac{dx_{j}}{\sqrt{2 i
\pi}}\ .
\end{equation}
Then the limit $F^{(sc)} = \lim_{n \to \infty} F_n^{(sc)}$ is
defined as a path distribution as above and the \textit{scattering
matrix} is defined by
\begin{eqnarray} S(k_{out},k_{in}) &=& \bigg\langle \exp \left[- i
\sum_{j=-\infty}^\infty V \big(x_j + x_0 + k_{in} (j\wedge 0) +
k_{out}(j \vee 0) \big) \lambda_j \right] \nonumber \\ && \qquad
\qquad \times e^{- i (k_{out}-k_{in}) x_0 },\, F^{(sc)}
\bigg\rangle.
\end{eqnarray} In this case of course we must take $\lambda_j =
e^{-\eps |j|}$. If $V$ decays sufficiently fast for $|x| \to
+\infty$, it is known that the limit $\eps \to 0$ exists.

\section{Appendix}

Here we give a proof of Sazonov's theorem based on \cite{SF}. 
We use a special case of Prokhorov's theorem \cite{Bourbaki} (see also \cite{Thomas2}):

\begin{theorem}[Prokhorov] Consider a separable Hilbert space
$\cal H$ with orthonormal basis $(e_n)_{n \in \NN}$, and let
$(\mu_N)_{N \in \NN}$ be a projective sequence of (in general
complex-valued) measures on ${\cal H}$. Assume that $\sup_{N \in \NN} || \mu_N|| < +\infty$, and that for all $\eps > 0$, 
there exists a weakly compact set $K \subset {\cal H}$ such that
$$ |\mu_N| (\pi_N(K)^c) < \eps \quad \forall N \in \NN. $$
(Here $\pi_N(K)^c$ denotes the complement of $\pi_N(K)$ in $\pi_N(\cal{H})$.)
Then there exists a bounded projective limit measure  $\mu =
\varprojlim \mu_N$ on ${\cal H}_\sigma$ such that $\mu_N =
\pi_N(\mu)$.
\end{theorem}

This theorem is proved by remarking that ${\cal H}_\sigma$ is a completely regular topological space and therefore has 
a Stone-\v Cech compactification. Using the Riesz-Markov theorem, it then
follows that it suffices to define $$ \langle F,\,\mu\rangle =
\int_{\cal H} F\,d\mu $$ for all bounded continuous functions $F$
on ${\cal H}_\sigma$. One then defines $\langle F,\, \mu\rangle =
\lim_{N \to \infty} \int F \circ j_N \, d\mu_n$, where $j_N: {\cal
H}^{(N)} \to {\cal H}$ is the canonical inclusion of the span
${\cal H}^{(N)}$ of $\{e_1,\dots,e_N\}$ into $\cal H$. The limit
exists by Prokhorov's condition and the uniform continuity of $F$
on compacta as in \cite{Dorlas}.

We also use a simple lemma from \cite{SF}:

\begin{lemma} Let $\nu$ be a probability measure on $\RR^N$ and
assume that for a given $\eps > 0$, there exists a positive
definite $N \times N$ matrix $A$ such that
$$ \langle \xi,\,A  \xi \rangle \leq 1 \implies \left| 1- \int
e^{i \langle \xi, x \rangle} \nu(dx) \right| \leq \eps. $$ Then,
for all $R > 0$,
$$ \nu \left( B_N(R)^c \right) \leq c
\left( \eps + \frac{2}{R^2} \Tr(A) \right), $$ where $c > 0$ is an
absolute constant and $B_N(R) = \{x\in \RR^N:\, ||x|| \geq R\}$.
\end{lemma}

\textit{Proof.} Since for $||x|| \geq R$, $e^{-||x||^2/(2R^2)}
\leq e^{-1/2}$, we have
\begin{eqnarray*} && (1-e^{-1/2}) \nu(B_N(R)^c) \\ && \qquad \leq
\int_{\RR^N} \left( 1-e^{-||x||^2/(2 R^2)} \right) \nu(dx) \\ &&
\qquad = \int \frac{d\xi}{(2 \pi R^{-2})^{N/2}} e^{-R^2
||\xi||^2/2} \int (1-e^{i\langle \xi,\, x\rangle}) \nu(dx) \\
&& \qquad = \int 1_{\{\xi:\, \langle \xi,\, A\xi\rangle \leq 1\}}
\dots + \int 1_{\{\xi:\, \langle \xi,\, A\xi\rangle > 1\}} \dots
\\ && \qquad \leq \eps \int \frac{d\xi}{(2 \pi R^{-2})^{N/2}}
e^{-R^2 ||\xi||^2/2} + 2 \int_{\{\xi:\,\langle
\xi,\, A \xi\rangle > 1\}} \frac{d\xi}{(2 \pi R^{-2})^{N/2}} e^{-R^2 ||\xi||^2/2} \\
&& \qquad \leq \eps + 2 \int \sum_{n,m=1}^N \xi_n A_{nm}
\xi_m  e^{-R^2 ||\xi||^2/2} \frac{d\xi}{(2 \pi R^{-2})^{N/2}} \\
&& \qquad = \eps + 2 \sum_{n=1}^N A_{nn} \int \xi_n^2 e^{-R^2
||\xi||^2/2} \frac{d\xi}{(2 \pi R^{-2})^{N/2}} = \eps +
\frac{2}{R^2} \Tr(A). \end{eqnarray*} \phantom{xx} \hfill
$\square$

\textit{Proof of Sazonov's theorem}  Let $\eps > 0$. By
Prokhorov's theorem, we need to prove that there exists a (weakly)
compact set $K \subset {\cal H}'$ such that $|\mu_N| (\pi'_N(K)^c)
< \eps$ for all $N$. Given $\eta > 0$, there exists a
Hilbert-Schmidt map $u$ on $\cal H$ such that
$$ ||u \xi || \leq 1 \implies |\Phi_N(\xi) - \Phi_N(0)| < \eta, $$
where $$ \Phi_N(\xi) = \int e^{i \langle \pi_N(\xi), \, x\rangle}
\nu_N(dx). $$  Set
$$ K = \{ x \in {\cal H}':\, ||x|| \leq R \}. $$ This set is weakly
compact by the Banach-Alaoglu theorem. Now, if $\xi \in {\cal
H}^{(N)}$, the span of $\{e_1,\dots,e_N\}$, then $$ || u \xi ||^2
= \sum_{n=1}^\infty \sum_{m,m'=1}^N u_{nm} u_{nm'} \xi_m \xi_{m'}
= \langle \xi,\, A\xi \rangle $$ where $$ A_{mm'} =
\sum_{n=1}^\infty u_{nm} u_{nm'} = \langle e_m,\, u^T u (e_{m'}
\rangle. $$ Hence $A \geq 0$ and by the lemma applied to the
probability measure $\overline{\nu}_N = \nu_N/||\nu_N||$,
$$ \nu_N(B_N(R)^c) \leq c\,||\nu_N||\, (\eta + \frac{2}{R^2} \Tr(A))
\leq c\,||\nu_N||\, (\eta + \frac{2}{R^2} ||u||_{HS}^2). $$ Taking
$R = ||u||_{HS} \sqrt{2} \eta$ and $\eta = \eps/(2c \sup_{N \in
\NN}||\nu_N||)$, we have, since $\pi_N(K) = B_N(R)$,
$|\mu^{(N)}|(\pi_N(K)^c) \leq \eps.$  $ \square $

\end{document}